\documentclass[runningheads,a4paper]{llncs}

\usepackage{amsmath,amssymb}
\usepackage{tikz}
\usepackage{enumitem}
\usepackage{showexpl}
\usepackage{multirow}
\usepackage{wrapfig}
\usepackage{subcaption}
\usepackage{algorithm}
\usepackage{algpseudocode}
\usetikzlibrary{automata,shapes.multipart}
\usetikzlibrary[arrows,decorations.pathmorphing,backgrounds,positioning,fit,petri]

\begin{document}
\mainmatter
\title{Towards the Existential Control of Boolean Networks: A Preliminary Report}
\subtitle{(Extended Abstract)}
\titlerunning{Existential Control of Boolean Networks}

\author{Soumya Paul\inst{1} \and Jun Pang\inst{1,2} \and Cui Su\inst{1}}
\authorrunning{Paul et al.}

\institute{
Interdisciplinary Centre for Security, Reliability and Trust
\and
Faculty of Science, Technology and Communication,
University of Luxembourg\\
\email{firstname.lastname@uni.lu}
}

\maketitle
\newcommand{\bn}{{\sf BN}}
\newcommand{\Attr}{\mathcal{A}}
\newcommand{\naturals}{\mathbb{N}}
\newcommand{\graph}{\mathcal{G}}
\newcommand{\state}{{\bf s}}
\newcommand{\tstate}{{\bf t}}
\newcommand{\update}{{\xi}}
\newcommand{\St}{{\bf S}}
\newcommand{\T}{{\bf T}}
\newcommand{\x}{{\bf x}}
\newcommand{\f}{{\bf f}}
\newcommand{\tssyn}{{\sf TS}^{\sf sn}}
\newcommand{\tsasyn}{{\sf TS}^{\sf asn}}
\renewcommand{\ts}{{\sf TS}}
\newcommand{\tshat}{\overline{\sf TS}}
\newcommand{\reach}{{\sf reach}}
\newcommand{\scc}{{\sf SCC}}
\newcommand{\trs}{{\sf TRS}}
\newcommand{\prt}{{\sf par}}
\newcommand{\blocks}{\mathcal{B}}
\newcommand{\attr}{\mathcal{A}}
\newcommand{\B}{\overline{B}}
\newcommand{\cross}{\otimes}
\newcommand{\ats}{{\sf ATS}}
\newcommand{\bts}{{\sf BTS}}
\newcommand{\apts}{{\sf APTS}}
\newcommand{\ctr}{{\sf ctr}}
\newcommand{\control}{{\sf C}}
\newcommand{\cset}{{\mathbb{C}}}
\newcommand{\abc}{{\sf abc}}
\newcommand{\atsasyn}{{\sf ATS}^{\sf asn}}
\newcommand{\atssyn}{{\sf ATS}^{\sf sn}}
\newcommand{\btsasyn}{{\sf BTS}^{\sf asn}}
\newcommand{\btssyn}{{\sf BTS}^{\sf sn}}
\newcommand{\ac}{{\sf ac}}
\newcommand{\bas}{{\sf bas}}
\newcommand{\SB}{{\sf SB}}
\newcommand{\WB}{{\sf WB}}
\newcommand{\wb}{{\sf WB}}
\newcommand{\hd}{{\sf hd}}
\newcommand{\bigo}{\mathcal{O}}
\newcommand{\move}[1]{\stackrel{#1}{\longrightarrow}}
\newcommand{\path}{\rho}
\newcommand{\pre}{{\sf pre}}
\newcommand{\post}{{\sf post}}
\newcommand{\mat}{{\sf M}}
\newcommand{\lat}{\mathcal{L}}
\newcommand{\lbl}{\ell}
\newcommand{\ind}{{\sf ind}}

\newtheorem{observation}{Observation}

\begin{abstract}
  Given a Boolean network $\bn$ and a subset $\Attr$ of attractors of $\bn$, we study the problem of identifying a minimal subset $\control_\bn$ of
  vertices of $\bn$, such that the dynamics of $\bn$
  can reach from a state $\state$ in any attractor $A_s\in \Attr$ to any attractor $A_t\in \Attr$ by controlling (toggling) a subset of vertices in $\control_\bn$ in a single
  time step. We describe a method based on the decomposition of the network structure into strongly connected components called
  `blocks'. The control subset can be locally computed for each such block and the results then merged to derive the
  global control subset $\control_\bn$. This potentially improves the efficiency for many real-life networks that are large but
  modular and well-structured. We are currently in the process of implementing our method in software.
\end{abstract}

\section{Introduction}\label{sec:intro}
%
Systems biology, with the help of mathematical modelling, has revolutionised the human diseasome research
and paved the way towards the development of new therapeutic approaches and personalised medicine. Such therapies
target specific proteins within the cellular systems aiming to drive it from a `diseased' state to a `healthy' state.
However, it has been observed that disease-networks are intrinsically robust against perturbations due to the
inherent diversity and redundancy of compensatory signalling pathways~\cite{Hop08}. This greatly reduces the
efficacy of single-target drugs. Hence, rather than trying to design selective ligands that target individual
receptors only, network polypharmacology seeks to modify multiple cellular targets to tackle the compensatory
mechanisms and robustness of disease-associated cellular systems. This motivates the question of identifying
multiple drug targets using which the network can be `fully
controlled', i.e. driven from any (diseased) state to any desired target (healthy) state. Furthermore, for the
feasibility of the synthesis of such drugs, the number
of such targets should be minimised. However, biological networks are intrinsically large (number of components, parameters,
interactions, etc.) which results in an exponentially increasing number of potential drug target combination
making a purely experimental approach quickly infeasible. This reinforces the need of mathematical modelling and
computational techniques.

Boolean networks (BNs), first introduced by Kauffman~\cite{KS69}, is a popular and well-established
framework for modelling gene regulatory networks (GRNs) and their associated signalling pathways.
Its main advantage is that it is simple
and yet able to capture the important dynamical properties of the system under study, thus
facilitating the modelling of large biological systems as a whole. The states of a BN are tuples of 0s
and 1s where each element of the tuple represents the level of activity of a particular protein in the
GRN or the signalling pathway it models - 0 for inactive and 1 for active.
The BN is assumed to evolve dynamically by moving from one state to the next governed by a Boolean
function for each of its components. The steady state behaviour of a BN is given by its subset of states
called {\em attractors} to one of which the dynamics eventually settles down. In biological context,
attractors are hypothesised to characterise cellular phenotypes~\cite{KS69} and also correspond to
functional cellular states such as proliferation, apoptosis, differentiation, etc.~\cite{HS01}. The {\em control}
of a BN therefore refers to the reprogramming/changing of the parameters of the BN (functions, values of variables, etc.)
so that its dynamics eventually reaches a desired attractor or steady state.

The full control of linear networks is a well-studied problem~\cite{Kal63} and such control strategies have been
proposed over the years. Recent work on network controllability has shown that full controllability and
reprogramming of intercellular networks can be achieved by a minimum number of control targets~\cite{LSB11}.
However, the full control of non-linear networks is apparently more challenging predominantly due to the explosion of the
potential search space with the increase in the network size. There has not been a lot of work in this regard.
Kim et al.~\cite{KSK13} developed a method to identify the so-called `control kernel' which is a minimal set of nodes for
fully controlling a biological network. But, their method is based on the construction of the full state
transition graph of the network and as such does not scale well for large networks.

The BNs used to model real-life biological networks have multiple attractors, the sizes and distribution of which
are governed by certain power laws~\cite{GB09}. However in most of the cases only some of these
attractors are `biologically relevant', i.e. correspond to meaningful expressions of the GRNs.
Thus, focussing on only the relevant attractors might help reduce the complexity of the
control problem while still being biologically meaningful.

\smallskip\noindent
{\bf Our contributions.} In this work, we report the initial results on  a method for the control
of Boolean networks that exploits both their structural and
dynamic properties, as shown inevitable in~\cite{GR16}.
More precisely, given a Boolean network $\bn$ and a set of `relevant' attractors $\Attr$ of $\bn$,
the method computes a minimal set of variables (the {\em minimal control set}),
such that starting from an initial attractor $A_s\in \Attr$ and by controlling specific subsets of these variables in a {\em single time-step},
the BN can (potentially) reach any desired target attractor $A_t\in\Attr$  when left to evolve on its own according to its original
dynamics. A welcome side-effect of the method is that when $\Attr$ is the set of all attractors of $\bn$, it gives the minimal
set of vertices for fully controlling $\bn$. We use an approach that we have developed for the problem of target control (driving the BN to a given single target attractor)
of BNs, based on the decomposition of its network structure into strongly connected components called `blocks'.
Although the method can be applied on the entire BN in one-go, we believe that using the decomposition-based approach
can greatly increase its efficiency on large real-life biological networks whose BN models have well-behaved
modular structure. This is work in progress and we are currently implementing our method in software
to test its effectiveness on various networks.

\section{Background and Notations}\label{sec:background}
%
Let $N=\{1,2,\ldots, n\}$ where $n\geq 1$. A {\em Boolean network} is a tuple $\bn = (\x,\f)$ where $\x=(x_1,x_2,\ldots, x_n)$ such that each $x_i$ is a Boolean variable and
$\f=(f_1,f_2,\ldots,f_n)$ is a tuple of Boolean functions over $\x$. In what follows, $i$ will always range over $N$, unless stated otherwise.
A Boolean network $\bn=(\x,\f)$ may be viewed as a directed graph $\graph_\bn = (V,E)$, where $V=\{v_1,v_2\ldots, v_n\}$ is the set of {\em vertices}
or {\em nodes} (intuitively, $v_i$ corresponds to the variable $x_i$ for all $i$) and for every $i,j\in N$, there is a directed edge from $v_j$ to $v_i$, often denoted as $v_j\rightarrow v_i$, if and only if
$f_i$ depends on $x_j$. Thus $V$ is ordered according to the ordering of $\x$. For any vertex $v_i\in V$, we let $\ind(v_i)=i$ be the index of $v_i$ in this ordering.
For any subset $W$ of $V$, $\ind(W) = \{\ind(v)|\ v\in W\}$.
A {\em path} from a vertex $v$ to a vertex $v'$ is a (possibly empty) sequence of edges from $v$ to $v'$ in $\graph_\bn$.
For any vertex $v\in V$ we define its set of {\em parents} as $\prt(v)=\{v'\in V\ |\ v'\rightarrow v\}$ and for any subset $W$ of $V$,
$\prt(W) = \{\prt(v)\ |\ v\in W\}$. For the rest of the exposition, we assume
that an arbitrary but fixed network $\bn$ of $n$ variables is given to us and $\graph_\bn=(V,E)$ is its associated directed graph.

A {\em state} $\state$ of $\bn$ is an element in $\{0,1\}^n$. Let $\St$ be the set of states of $\bn$. For any state $\state=(s_1,s_2,\ldots,s_n)$, and
for every $i$, the value of $s_i$, often denoted as $\state[i]$, represents the value that the variable $x_i$ takes when the $\bn$ `is in state $\state$'.
For some $i$, suppose $f_i$ depends on $x_{i_1},x_{i_2},\ldots, x_{i_k}$. Then $f_i(\state)$ will denote the value $f_i(\state[i_1],\state[i_2],\ldots, \state[i_k])$.
For two states $\state,\state'\in\St$, the {\em Hamming distance} between $\state$ and $\state'$ will be denoted as $\hd(\state,\state')$ and $\arg(\hd(\state,\state'))\subseteq N$ will denote the set of indices in which $\state$ and $\state'$ differ. For a state
$\state$ and a subset $\St'\subseteq\St$, the Hamming distance between $\state$ and $\St'$ is defined as
$\hd(\state,\St')=\min_{\state'\in\St'}\hd(\state,\state')$. We let $\arg(\hd(\state,\St'))$ denote the set of subsets of $N$ such
that $I\in \arg(\hd(\state,\St'))$ if and only if $I$ is a set of indices of the variables that realise $\hd(\state,\St')$. 

We assume that the Boolean network starts initially in a state $\state_0$ and its state changes in every discrete time-step according to the update
functions $\f$. In this work, we shall deal with the asynchronous updating scheme but all our results transfer to the synchronous updating scheme as well.
Suppose $\state_0\in\St$ is an initial state of $\bn$. The {\em asynchronous evolution} of $\bn$ is a function $\update: \naturals \rightarrow \wp(\St)$
such that $\update(0)=\state_0$ and for every $j\geq 0$, if $\state\in\update(j)$ then $\state'\in \update(j+1)$ if and only if  either $\hd(\state,\state') = 1$
and $\state'[i]=f_i(\state)$ where $i=\arg(\hd(\state,\state'))$ or $\hd(\state,\state')=0$ and there exists $i$ such that $\state'[i]=f_i(\state)$.

The dynamics of a Boolean network can be represented as a {\em state transition graph} or a {\em transition system (TS)}. The {\em transition system}
of $\bn$, denoted as $\ts_\bn$ is a tuple $(\St,\rightarrow)$ where the vertices are the set of states $\St$ and for any two states
$\state$ and $\state'$ there is a directed edge from $\state$ to $\state'$, denoted $\state\rightarrow\state'$, if and only if  either $\hd(\state,\state') = 1$
and $\state'[i]=f_i(\state)$ where $i=\arg(\hd(\state,\state'))$ or $\hd(\state,\state')=0$ and there exists $i$ such that $\state'[i]=f_i(\state)$.
\vspace{-0.2mm}

For any state $\state\in \St$, $\pre_\ts(\state) = \{\state'\in \St\ |\ \state'\rightarrow \state\}$  contains all the states that can
reach $\state$ by performing a single transition in $\ts$. For a subset $\St'$ of $\St$, $\pre_\ts(\St') = \bigcup_{\state\in\St'}\pre_\ts(\state)$.
A {\em path} from a state $\state$ to a state $\state'$ is a (possibly empty) sequence of transitions from $\state$ to $\state'$ in $\ts_\bn$. A path from a
state $\state$ to a subset $\St'$ of $\St$ is a path from $\state$ to any state $\state'\in \St'$.
For a state $\state\in\St$, $\reach_\ts(\state)$ denotes the set of states $\state'$ such that there is a path from $\state$ to $\state'$ in $\ts$.

An {\em attractor} $A$ of $\ts_\bn$ (or of $\bn$) is a subset of states of $\St$ such that for every $\state\in A,\ \reach_{\ts_\bn}(\state)=A$.
Any state which is not part of an attractor is a {\em transient state}. An attractor $A$ of $\bn$ is said to be reachable from a state $\state$ if
$\reach_{\ts_\bn}(\state)\cap A\neq\emptyset$. Attractors represent the stable behaviour of the $\bn$ according to the dynamics. For an attractor $A$ of $\bn$,
the {\em weak basin} or simply the {\em basin} of attraction of $A$, denoted $\bas_{\ts_\bn}(A)$, is a subset of states of $\St$ such that $\state\in\bas_{\ts_\bn}(A)$ if
$\reach_{\ts_\bn}(\state)\cap A\neq \emptyset$. A {\em control} $\control$ is a (possibly empty)
subset of $N$. For a state $\state\in \St$, the {\em application of control} $\control$ to $\state$,
denoted $\control(\state)$ is defined as the state $\state'\in \St$ such that $\state'[i]=(1-\state[i])$ if $i\in \control$ and $\state'[i]=\state[i]$,
otherwise. Henceforth, we drop the subscripts $\ts$ or $\bn$ or both when no ambiguity arises.

\smallskip
\noindent {\bf Control problems:}  In this work we shall exclusively deal with the notion of {\em existential control}
in that, after the control $\control$ is applied to a state $\state$, there `exists' a path from $\control(\state)$ to the desired target attractor and also
perhaps to other non-target attractors. This is different from the notion of {\em absolute control} dealt with in~\cite{PSPM18} where after the control, $\control(\state)$ is
`guaranteed' to reach the target attractor. Although the techniques applied for the computation of the minimal control are similar in both cases,
there are certain fundamental differences. In particular, here we are interested
in the following control problems given a network $\bn$. Note that for us, the control is applied in a single time step (hence simultaneously) to the state
$\state$ under consideration.
\begin{enumerate}
\item {\bf Minimal existential target control:} Given a state $\state\in\St$ and a `target attractor' $A_t$ of $\bn$, it is a control $\control_{\state\rightarrow A_t}$
  such that after the application of $\control_{\state\rightarrow A_t}(\state)$, $\bn$ can eventually reach $A_t$ and $\control_{\state\rightarrow A_t}$ is a minimal
  such subset.
\item {\bf Minimal existential all-pairs control:} Given a set $\attr=\{A_1,A_2,\ldots, A_p\}$, $p\geq 2$, of attractors of $\bn$, it is a minimal subset $\control_\attr$ of
  $N$ such that for any pair $A_i,A_j \in \attr$ of attractors, there is a state $\state\in A_i$, such that $\control_{\state\rightarrow A_j} \subseteq \control_\attr$.
  \item {\bf Minimal existential full control:} $\control_\bn$ is the minimal existential all-pairs control $\control_\attr$ when $\attr$ is the set of all attractors of $\bn$.
\end{enumerate}

In this work we shall use ideas from the decomposition-based approach of~\cite{PSPM18} to compute (2) and (3). We first give the relevant definitions and results.

Let $\scc$ denote the set of maximal strongly connected components (SCCs) of $\graph_\bn$.
A {\em basic block} $B$ is a subset
of nodes of $\bn$ such that $B=(S\cup\prt(S))$ where $S$ is a maximal SCC of $\graph_\bn$. Let $\blocks$ denote the set of basic blocks of $\bn$. The union
of two or more basic blocks will also be called a {\em block}.
Using the set of basic blocks as vertices, we can form a directed graph $\graph_\blocks = (\blocks, E_\blocks)$, called the {\em block graph} of $\bn$,
where for any pair of basic blocks $B',B\in \blocks, B'\neq B$, there is a directed edge from $B'$ to $B$ if and only if $B'\cap B\neq\emptyset$ and
for every $v\in B'\cap B$, $\prt(v)\cap B =\emptyset$. In such a case, $B'$ is called a {\em parent} block of $B$ and $v$ is called a {\em control node}
for $B$. The set of parent blocks of $B$ is denoted as $\prt(B)$.

A block is called {\em elementary} if $\prt(B)=\emptyset$ and {\em non-elementary} otherwise. We shall henceforth assume that $\bn$ has $k$ basic blocks,
$|\blocks|=k$, and $\graph_\bn$ is topologically sorted as $\{B_1,B_2,\ldots,B_k\}$. Given how $\graph_\bn$ is constructed, it will be a directed acyclic graph
and hence can always be topologically sorted. Note that for every $j:1\leq j\leq k$, $(\bigcup_{\ell=1}^j B_\ell)$ is an
elementary block. We shall denote it as $\B_j$ and let $B_j^-=(B_j\setminus\B_{j-1})$. For two basic blocks $B$ and $B'$ where $B$ is non-elementary, $B'$
is said to be an {\em ancestor} of $B$
if there is a path from $B'$ to $B$ in the block graph $\graph_\blocks$. The {\em ancestor-closure} of a basic block $B$, 
denoted $\ac(B)$ is defined as the union of $B$ with all its ancestors. Note that $\ac(B)$ is an elementary block and so is $(\ac(B)\setminus B^-)$,
denoted as $\ac(B)^-$.

For a block $B$ of $\bn$,
its state space is $\{0,1\}^{|B|}$ and is denoted as $\St_B$. For any state $\state\in \St$, where $\state=(s_1,s_2,\ldots, s_n)$, the projection of $\state$
to $B$, denoted $\state|_B$ is the tuple obtained from $\state$ by suppressing the values of the variables not in $B$. Let $B_1$ and $B_2$ be two blocks of
$\bn$ and let $\state_1$ and $\state_2$ be states of $B_1$ and $B_2$, respectively. $\state_1\cross\state_2$ is defined (called {\em crossable}) if there exists a
state $\state\in \St_{B_1\cup B_2}$ such that $\state|_{B_1}=\state_1$ and $\state|_{B_2}=\state_2$. $\state_1\cross\state_2$ is then defined to be this unique
state $\state$. For any subsets $\St_1$ and $\St_2$ of $\St_{B_1}$ and $\St_{B_2}$ resp. $\St_1\cross\St_2$ is a subset of $\St_{B_1\cup B_2}$ and is defined as: $\St_1\cross\St_2 = \{\state_1\cross\state_2\ |\ \state_1\in\St_1, \state_2\in\St_2 \text{ and } \state_1,\state_2 \text{ are crossable}\}$.
The cross operation can be defined for more than two states $\state_1,\state_2,\ldots,\state_k$, as $\state_1\cross\state_2\cross\ldots\state_k=(((\state_1\cross\state_2)\cross\ldots)\cross\state_k)$. The cross operation can be similarly lifted to more than two sets of states.

The TS $\ts_B$ of an elementary block $B$ of $\bn$ is defined similarly to the TS of $\bn$, which can indeed be done as
the update functions do not depend on vertices outside $B$. The attractors, basin of attractions, etc. of such a TS is also defined similarly.
The TSs of a non-elementary basic block $B$ are `realised' by the basins of attractions of the attractors
of $\ac(B)^-$, each such attractor realising a different TS. Thus, if $A$ is an attractor of $\ac(B)^-$ then $\ts_B$
realised by $\bas(A)$ has set of states $\St$ which the maximum subset of $\St_{\ac(B)}$ such that $\St|_{\ac(B)^-}= \bas(A)$. The transitions are then defined as
usual. The following is a key result, a counterpart of which
was proved in~\cite{PSPM18}, saying that the `global' attractors of $\bn$ and their basins can be computed by first computing the `local' attractors
and basins of the basic blocks and then merging them using the cross operation.

\begin{theorem}[\cite{PSPM18}]\label{thm:pres} $A$ is an attractor of $\bn$ iff there exist attractors $A_j$ of $B_j$ such that $A_j=A|_{B_j}$ for all $j:1\leq j\leq k$ and 
  $A = \cross_{j}A_j$. Furthermore, $A|_{\ac(B_j)}$ is an attractor of $\ac(B_j)$ and $\bas(A) = \cross_j \bas(A_j)$ w.r.t. their TSs.
\end{theorem}

\section{Results}\label{sec:results}
%
In this section we develop our method for solving control problem (2). We first describe a `global' approach that works on the entire BN
and then modify it to exploit the decomposition-based approach of~\cite{PSPM18}. For simplicity, we assume that every attractor of
$\bn$ is a single state with a self loop. The methods can be generalised for the case where an attractor can comprise of two or
more states.

First, note that given a state $\state$ and an attractor $A$, for $\bn$ to potentially end up in $A$ after the application of a
control $\control$, it is necessary and sufficient that there is a path from $\control(\state)$ to $A$ in $\ts_\bn$ which means, by definition,
that $\control(\state)\in\bas(A)$. Thus given a set $\attr$ of attractors of $\bn$ to compute $\control_\attr$ it is enough to compute the basins of
the attractors in $\attr$. This can be done, starting from any attractor in $\attr$, by a repeated application of the $\pre(\cdot)$ operator till a
fixed point is reached. The procedure {\sc Compute\_Basin} described in Algorithm~\ref{alg:basin} does exactly this.

\begin{algorithm}[!t]
\caption{Computation of the basin of attraction}
\label{alg:basin}
\begin{algorithmic}[1]
  \Procedure{Compute\_Basin}{$\f,A$}
  \State {\sc Bas} = $A$
  \State Till {\pre({\sc Bas})$\neq$ {\sc Bas}} do
  \State \qquad {\sc Bas} = \pre({\sc Bas})
  \State done
  \State \Return {\sc Bas}
  \EndProcedure
\end{algorithmic}
\end{algorithm}

So, assume that the given set of attractors $\attr$
is sorted as $\{A_1, A_2,\ldots, A_p\}$. We then construct a $p\times p$ matrix $\mat$ whose entries are subsets of $N$ and
are defined as: for every $I\subseteq N$, $I\in \mat_{ij}$ if and only if $I = \arg(\hd(\state,\state'))$ where $\state\in A_i$
and $\state'\in\bas(A_j)$. That is, for every pair of attractors $A_i$ and $A_j$ the entries of $\mat_{ij}$ record the indices of the variables
that need to be toggled in state $\state\in A_i$ to end up in any of the states of the basin of $A_j$. The minimal all-pairs control
$\control_\attr$ is then nothing but a minimal subset of $N$ such that for every $i,j$ there exists $I\in\mat_{ij}$ such that
$I\subseteq \control_\attr$.

We now describe a method to compute the set $\control_\attr$ based on the power-set lattice of $N$, denoted by $\lat$. Let
$\lbl:\lat \rightarrow \wp(N \times N)$ be a labelling function that labels the elements of $\lat$ with tuples in
$(N\times N)$ defined as follows. For any element $L$ of $\lat,\ (i,j)\in \lbl(L)$ iff $L\in \mat_{ij}$. Let $\lbl^*$
denote the closure of the labelling function of $\lat$ under subsets, defined as: for every element $L$ of $\lat$, $\lbl^*(L) = \bigcup_{L'\subseteq L}\lbl(L')$.
Finally, $\control_\attr$ is any minimal element $L$ of $\lat$ such that $\lbl^*(L) = (\{1,2,\ldots,p\}\times  \{1,2,\ldots,p\})\setminus \{(i,i)\ |\ i\in\{1,2,\ldots, p\}\}$.
Control problem (3) is a special case of (2) where $\attr$ is the set of all attractors of $\bn$. For solving (3), given a $\bn$ as input,
we can first apply any of the methods available in the literature (e.g., see~\cite{MPQY17,MPQY17b}) to compute the set of all attractors $\attr$ of $\bn$,
and then invoke the above method.

In general, the problem of computing $\control_\attr$ given the matrix $\mat$ is NP-hard. Moreover, given a BN and an attractor $A$ as input,
the problem of computation of the strong basin of $A$ is PSPACE-hard. Hence, the control problem (2) is at least PSPACE-hard and so unlikely
to have efficient algorithms for the general case. However, in~\cite{PSPM18} we show that using a decomposition-based approach we can improve the efficiency
for many modular well-structured networks. We now describe a similar approach for solving control problem (2) [and hence (3)].

The method is iterative where instead of computing the basin of attractions of the given attractors for the entire BN in one-go, we decompose the BN into
blocks, as described in the previous section, and compute the basins and also the minimal control w.r.t the transition system of each such block. The basin
of an attractor in a block can once again be computed using a repeated application of the $\pre(\cdot)$ operator in that block. The details
are given in Algorithm~\ref{alg:basindecomp}.

\begin{algorithm}[!t]
\caption{Decomposition-based computation of the basin of attraction}
\label{alg:basindecomp}
\begin{algorithmic}[1]
  \Procedure{Compute\_Basin\_Block}{$\f,A,B_k$,{\sc Bas\_Anc}}
  \State {\sc Bas} = $A$; {\sc Pre}=$\emptyset$
  \State do
  \State \qquad {\sc Pre} = {\sc Bas} $\cup$ \pre({\sc Bas})$\setminus$\{$\state\in$ \pre({\sc Bas})$\ |\ \state|_{\ac(B_k)^-}\notin$ {\sc Bas\_Anc}\}
  \State Till {\sc Pre} $\neq$ {\sc Bas}
  \State \Return {\sc Bas}
  \EndProcedure
\end{algorithmic}
\end{algorithm}

Suppose we are given a $\bn$ and a set of attractors $\attr$ sorted as $\{A_1,A_2,\ldots, A_p\}$ as input. We proceed in the following steps:
\begin{enumerate}
\item We decompose $\bn$ into basic blocks $\blocks$, form the block graph $\graph_\blocks$ and topologically sort it to obtain an ordering of
  the blocks as $\blocks=\{B_1,B_2,\ldots, B_k\}$.
\item Proceeding in the sorted order, for each block $B_j$ we repeat the steps below:
  \begin{enumerate}
    \item Let $\hat{B}_j=(B_j\setminus(\bigcup_{r<j}B_r))$ and $I_j = \ind(\hat{B}_j)$.
  \item Let $\mat^j$ be a $p\times p$ matrix whose entries are subsets of $I_j$.
  \item Note that by Theorem~\ref{thm:pres}, $A_r|_{\ac(B_j)}$ is an attractor of $B_j$, for every $r:1\leq r\leq p$. For every $r$, we compute
    $\bas(A_r|_{\ac(B_j)})$ using the basin of the parent block of $B_j$, $\bas(A_r|_{\ac(B_j)^-})$, by the application of the procedure {\sc Compute\_Basin\_Block}($\f,A_r,\blocks,\bas(A_r|_{\ac(B_j)^-})$) described in Algorithm \ref{alg:basindecomp}.
  \item We populate the matrix $\mat^j$ as: for every $I\subseteq I_j$, $I\in \mat^j_{qr}$ if and only if $I = (\arg(\hd(\state|_{\hat{B}_j},\state'|_{\hat{B}_j})))$
    for some $\state\in A_q|_{\ac(B_j)}$ and $\state'\in\bas(A_r|_{\ac(B_j)})$.
  \item Let $\lat_j$ be the subset lattice of $I_j$ and let $\lbl_j$ label the elements of $\lat_j$ with tuples in $(I_j\times I_j)$ such that for $L\in\lat_j,\ (q,r)\in \lbl_j(L)$ iff $L\in \mat^j_{qr}$.
  \item Let $\lbl^*_j$ denote the closure of $\lbl_j$ under subsets and let $\control^j_\attr$ be any minimal element $L$ of $\lat_j$ such that
    $\lbl^*(L) = ((\{1,2,\ldots,p\}\times  \{1,2,\ldots,p\})\setminus \{(i,i)\ |\ i\in\{1,2,\ldots, p\}\})$
  \end{enumerate}
  \item Finally we let $\control_\attr = \bigcup_{j=1}^k\control^j_\attr$.
\end{enumerate}

The above approach is worked-out in details on a toy example in the next section.

\section{A Detailed Example}\label{sec:example}
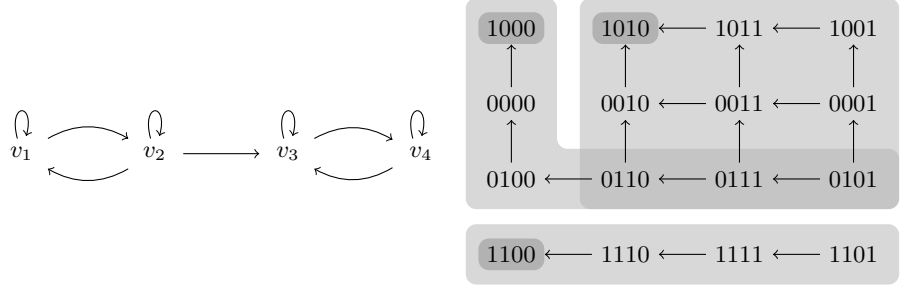
\begin{figure}[!t]
    \centering
    \begin{minipage}{0.5\textwidth}
\begin{tikzpicture}

\node        (s1)                  {$~v_1~$};
\node(s2)  [right=of s1]   {$~v_2~$};
\node        (s3)  [right=of s2]   {$~v_3~$};
\node (s4) [right=of s3] {$~v_4$};

 \path     
 (s1)  edge [<-,loop above] node {} (s1)
 (s1)  edge [->,bend left] node {} (s2)

 (s2)  edge [<-,loop above] node {} (s2)
 (s2)  edge [->,bend left] node {} (s1)
 (s2)  edge [->] node {} (s3)
 (s3)  edge [<-,loop above] node {} (s3)
 (s3) edge [->, bend left] node {} (s4)
 (s4) edge [->, bend left] node {} (s3)
 (s4)  edge [<-,loop above] node {} (s4)
 ;
\end{tikzpicture}
    \end{minipage}
    \begin{minipage}{0.45\textwidth}
{\begin{tikzpicture}
[attr/.style={draw=none,fill=black!30,rounded corners}]
\draw[draw=none,fill=black!30,rounded corners,opacity=0.5] (0,1.4) -- (0.6,1.4) -- (0.6,-0.6) -- (5.1,-0.6) -- (5.1,-1.4) -- (-0.6,-1.4) -- (-0.6,1.4) -- (0,1.4);
\draw[draw=none,fill=black!30,rounded corners,opacity=0.5] (1.5,1.4) -- (5.1,1.4) -- (5.1,-1.4) -- (0.9,-1.4) -- (0.9,1.4) -- (1.5,1.4);
\draw[draw=none,fill=black!30,rounded corners,opacity=0.5] (0,-1.6) -- (5.1,-1.6) -- (5.1,-2.4) -- (-0.6,-2.4) -- (-0.6,-1.6) -- (0,-1.6);
\node (t1)[attr] at (0,1) {1000};
\node (t2)[attr] at (1.5,1) {1010};
\node (t3) at (3,1) {1011};
\node (t4) at (4.5,1) {1001};
\node (t5) at (0,0) {0000};
\node (t6) at (1.5,0) {0010};
\node (t7) at (3,0) {0011};
\node (t8) at (4.5,0) {0001};
\node (t9) at (0,-1) {0100};
\node (t10) at (1.5,-1) {0110};
\node (t11) at (3,-1) {0111};
\node (t12) at (4.5,-1) {0101};
\node (t13)[attr] at (0,-2) {1100};
\node (t14) at (1.5,-2) {1110};
\node (t15) at (3,-2) {1111};
\node (t16) at (4.5,-2) {1101};

\path
(t3) edge [->] (t2)
(t4) edge [->] (t3)
(t5) edge [->] (t1)
(t6) edge [->] (t2)
(t7) edge [->] (t3)
(t7) edge [->] (t6)
(t8) edge [->] (t4)
(t8) edge [->] (t7)
(t9) edge [->] (t5)
(t10) edge [->] (t6)
(t10) edge [->] (t9)
(t11) edge [->] (t7)
(t11) edge [->] (t10)
(t12) edge [->] (t8)
(t12) edge [->] (t11)
(t14) edge [->] (t13)
(t15) edge [->] (t14)
(t16) edge [->] (t15)
;
\end{tikzpicture}}
    \end{minipage}
\caption{The graph of $\bn$ and its transition system.}
\label{fig:fullts}%
\end{figure}

%
Consider the four-node Boolean network $\bn=(\x,\f)$ where $\x=(x_1,x_2,x_3,x_4)$ and $\f=(f_1,f_2,f_3,f_4)$ where $f_1=\neg x_2 \lor
(x_1\land x_2), f_2=x_1\land x_2, f_3=x_4\lor(\neg x_2\land x_3)$ and $f_4=\neg x_3\land x_4$. The graph of the network $\graph_\bn$
and its associated transition system $\ts$ is given in Figure~\ref{fig:fullts}. Note that every state of $\ts$ also has a self loop (an edge
to itself) which we have not shown in the figure to avoid clutter, but is implicit. $\ts$ has three attractors $\attr=\{A_1,A_2,A_3\}$ shown
by dark grey rectangles, where $A_1=\{(1000)\}, A_2=\{(1100)\}$ and $A_3=\{(1010)\}$. Their corresponding basins of attractions are shown by
enclosing grey regions of a lighter shade. Note that the states $(0110),(0111)$ and $(0101)$ are in the basins of both the attractors $A_1$ and $A_3$.

 Now, suppose the relevant set of attractors given is $\attr=\{A_2,A_3\}$. We construct the $2\times 2$ matrix $\mat$ as given in Table~\ref{tab:matfull}, where the elements of $\mat_{A_2A_3}$
 are the subsets of $\{1,2,3,4\}$ needed to be controlled (toggled) in $A_2$ to end up in one of the states of the basin of $A_3$ and those of $\mat_{A_3A_2}$ are the subsets
 $\{1,2,3,4\}$ needed to be controlled in $A_3$ to move to the basin of $A_2$.

  \begin{table}[!h]
    \centering
    \begin{tabular}{|c||c|c|}
      \hline
           & $A_2$ & $A_3$ \\
      \hline
      \hline
      \multirow{2}{*}{$A_2$} & & \{1,3\}, \{1,4\}, \{2,3\}, \{2,4\}, \{1,2,3\}\\
      & & \{1,2,4\}, \{1,3,4\}, \{2,3,4\}, \{1,2,3,4\}\\
      \hline
      $A_3$ & \{2\},\{2,3\},\{2,4\},\{2,3,4\} &\\
      \hline
    \end{tabular}
    \caption{The $2\times2$ matrix $\mat$.}
    \label{tab:matfull}
    \vspace{-5mm}
  \end{table}

\begin{figure}
    \centering
    \begin{tikzpicture}
      \node (max) at (0,6) {$\{1,2,3,4\}$};
      \node [red,right] at (max.east) {\scriptsize{$(2,3)$}};
      \node (a) at (-3,4) {$\{1,2,3\}$};
      \node [red,left] at (a.west) {\scriptsize{$(2,3)$}};
      \node (b) at (-1,4) {$\{1,2,4\}$};
      \node [red,above] at (b.north) {\scriptsize{$(2,3)$}};
      \node (c) at (1,4) {$\{1,3,4\}$};
      \node [red,below] at (c.south) {\scriptsize{$(2,3)$}};
      \node (d) at (3,4) {$\{2,3,4\}$};
      \node [red,right] at (d.east) {\scriptsize{$(2,3),(3,2)$}};
      \node (e) at (-5,2) {$\{1,2\}$};
      \node (f) at (-3,2) {$\{1,3\}$};
      \node [red,above] at (f.north) {\scriptsize{$(2,3)$}};
      \node (g) at (-1,2) {$\{1,4\}$};
      \node [red,below] at (g.south) {\scriptsize{$(2,3)$}};
      \node (h) at (1,2) {$\{2,3\}$};
      \node [red,above] at (h.north) {\scriptsize{$(2,3),(3,2)$}};
      \node (i) at (3,2) {$\{2,4\}$};
      \node [red,below] at (i.south) {\scriptsize{$(2,3),(3,2)$}};
      \node (j) at (5,2) {$\{3,4\}$};
      \node (k) at (-3,0) {$\{1\}$};
      \node (l) at (-1,0) {$\{2\}$};
      \node [red,above] at (l.north) {\scriptsize{$(3,2)$}};
      \node (m) at (1,0) {$\{3\}$};
      \node (n) at (3,0) {$\{4\}$};
      \node (min) at (0,-2) {$\emptyset$};
      \draw (min) -- (k) -- (e) -- (a) -- (max) -- (b) -- (g)
      (l) -- (min)
      (min) -- (n) -- (j) -- (d) -- (max) -- (c) -- (g)
      (min) -- (m) -- (h) -- (d)
      (k) -- (f) -- (a)
      (n) -- (i) -- (d)
      (k) -- (g)
      (l) -- (h);
      \draw (l) -- (i)
      (l) -- (e)
      (m) -- (f)
      (m) -- (j)
      (n) -- (g)
      (e) -- (b)
      (f) -- (c)
      (h) -- (a)
      (i) -- (b)
      (j) -- (c);
    \end{tikzpicture}
    \caption{The subset lattice of $\{1,2,3,4\}$ with corresponding labels from $\mat$.}
    \label{fig:latfull}
  \end{figure}
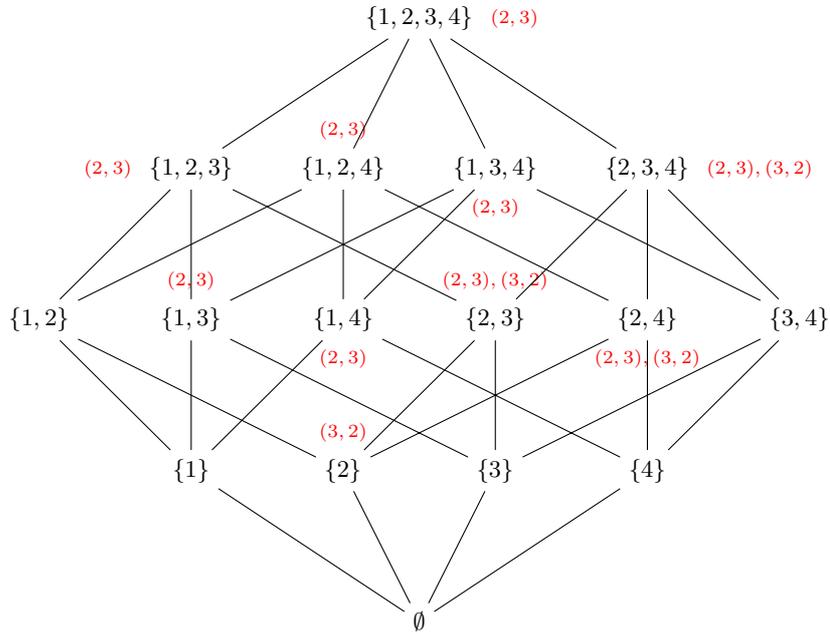

We next construct the subset lattice $\lat$ for \{1,2,3,4\} and label each element $L$ of $\lat$ with tuples in $(\{2,3\}\times\{2,3\})$ as:
$L$ is labelled with $(i,j)$ if and only if $L$ is an element of $\mat_{A_iA_j}$ (Figure~\ref{fig:latfull}).
  
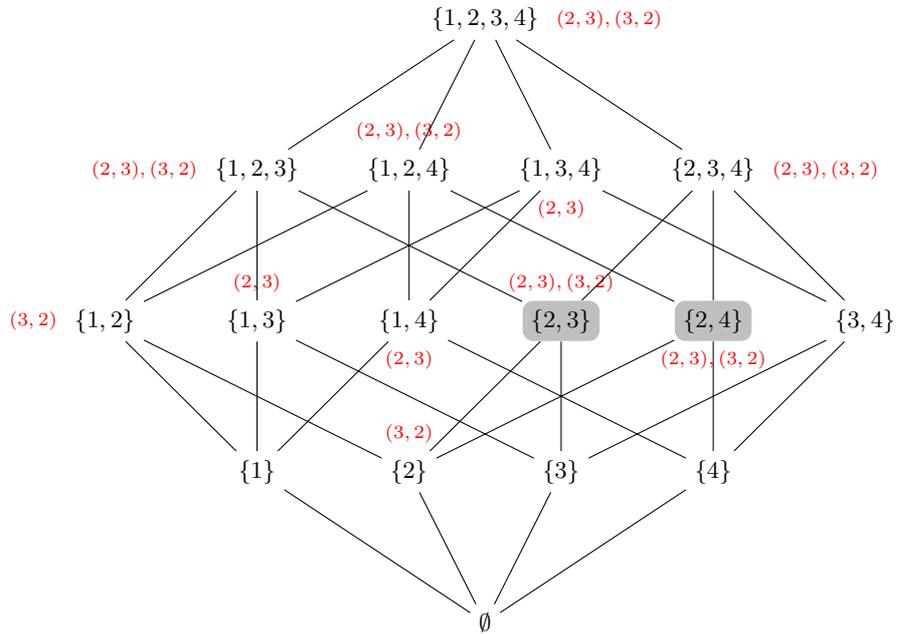
\begin{figure}
    \centering
    \begin{tikzpicture}
      \node (max) at (0,6) {$\{1,2,3,4\}$};
      \node [red,right] at (max.east) {\scriptsize{$(2,3),(3,2)$}};
      \node (a) at (-3,4) {$\{1,2,3\}$};
      \node [red,left] at (a.west) {\scriptsize{$(2,3),(3,2)$}};
      \node (b) at (-1,4) {$\{1,2,4\}$};
      \node [red,above] at (b.north) {\scriptsize{$(2,3),(3,2)$}};
      \node (c) at (1,4) {$\{1,3,4\}$};
      \node [red,below] at (c.south) {\scriptsize{$(2,3)$}};
      \node (d) at (3,4) {$\{2,3,4\}$};
      \node [red,right] at (d.east) {\scriptsize{$(2,3),(3,2)$}};
      \node (e) at (-5,2) {$\{1,2\}$};
      \node [red,left] at (e.west) {\scriptsize{$(3,2)$}};
      \node (f) at (-3,2) {$\{1,3\}$};
      \node [red,above] at (f.north) {\scriptsize{$(2,3)$}};
      \node (g) at (-1,2) {$\{1,4\}$};
      \node [red,below] at (g.south) {\scriptsize{$(2,3)$}};
      \node [draw=none,fill=black!25,rounded corners](h) at (1,2) {$\{2,3\}$};
      \node [red,above] at (h.north) {\scriptsize{$(2,3),(3,2)$}};
      \node [draw=none,fill=black!25,rounded corners](i) at (3,2) {$\{2,4\}$};
      \node [red,below] at (i.south) {\scriptsize{$(2,3),(3,2)$}};
      \node (j) at (5,2) {$\{3,4\}$};
      \node (k) at (-3,0) {$\{1\}$};
      \node (l) at (-1,0) {$\{2\}$};
      \node [red,above] at (l.north) {\scriptsize{$(3,2)$}};
      \node (m) at (1,0) {$\{3\}$};
      \node (n) at (3,0) {$\{4\}$};
      \node (min) at (0,-2) {$\emptyset$};
      \draw (min) -- (k) -- (e) -- (a) -- (max) -- (b) -- (g)
      (l) -- (min)
      (min) -- (n) -- (j) -- (d) -- (max) -- (c) -- (g)
      (min) -- (m) -- (h) -- (d)
      (k) -- (f) -- (a)
      (n) -- (i) -- (d)
      (k) -- (g)
      (l) -- (h);
      \draw (l) -- (i)
      (l) -- (e)
      (m) -- (f)
      (m) -- (j)
      (n) -- (g)
      (e) -- (b)
      (f) -- (c)
      (h) -- (a)
      (i) -- (b)
      (j) -- (c);
    \end{tikzpicture}
    \caption{The subset lattice of $\{1,2,3,4\}$ with labels closed under subsets.}
    \label{fig:latfullsubset}
  \end{figure}

  Next, we take the subset closure of the labels of the elements of $\lat$. That is, we label every element of $\lat$ with the label of itself
  and that of all its subsets. The resulting labelling of $\lat$ is shown in Figure~\ref{fig:latfullsubset}

We find that $\{2,3\}$ and $\{2,4\}$ are the minimal elements of $\lat$ whose label contains both $(2,3)$ and $(3,2)$, i.e., all
the pairs of the indices of the attractors in $\attr$. Hence, we conclude that $\control_\attr$ is equal to either $\{2,3\}$ or $\{2,4\}$.

However, as mentioned in Section~\ref{sec:results}, the problem of computing the set $\control_\attr$ from the matrix $\mat$ is NP-hard in general and the method based on the construction
of the subset lattice $\lat$ can be clearly exponential in $N$, the number of variables of the BN.
We can take advantage of the decomposition-based approach~\cite{MPQY17,MPQY17b,PSPM18} on certain
well-structured modular BNs by splitting it into `blocks' and performing the computations locally on the blocks. This can improve the efficiency for many real-life biological
networks whose BN models have such modular structures. We described the approach towards the end of Section~\ref{sec:results}. We now demonstrate it on our example BN.

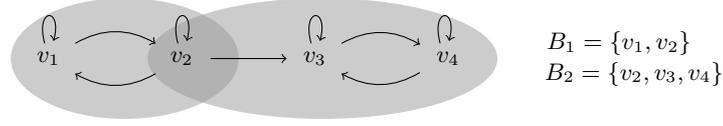
\begin{figure}[!t]
\centering
\begin{tikzpicture}
\draw[draw=none,fill=black!40,opacity=0.5] (1,0) ellipse (1.5cm and 0.8cm);
\draw[draw=none,fill=black!40,opacity=0.5] (3.6,0) ellipse (2.3cm and 0.8cm);

\node        (s1)                  {$~v_1~$};
\node(s2)  [right=of s1]   {$~v_2~$};
\node        (s3)  [right=of s2]   {$~v_3~$};
\node (s4) [right=of s3] {$~v_4$};

\node(b1) at (7.5,0.2) {$B_1=\{v_1,v_2\}$};
\node(b2) at (7.7,-0.2) {$B_2=\{v_2,v_3,v_4\}$};

 \path     
 (s1)  edge [<-,loop above] node {} (s1)
 (s1)  edge [->,bend left] node {} (s2)

 (s2)  edge [<-,loop above] node {} (s2)
 (s2)  edge [->,bend left] node {} (s1)
 (s2)  edge [->] node {} (s3)
 (s3)  edge [<-,loop above] node {} (s3)
  (s3) edge [->, bend left] node {} (s4)
 (s4) edge [->, bend left] node {} (s3)
 (s4)  edge [<-,loop above] node {} (s4)
 ;
\end{tikzpicture}
\caption{The blocks of $\bn$}
\label{fig:blocks}
    \end{figure}

Towards that, first note that $\bn$ has two SCCs $S_1=\{v_1,v_2\}$ and $S_2=\{v_3,v_4\}$. No node of $S_1$ has a parent node that does not belong to $S_1$. Hence $S_1$ forms
the elementary block $B_1=S_1$. Next, the parent of the node $v_3$ of $S_2$ is $v_2$. Hence $S_2$ forms the non-elementary block $B_2 = \{v_2,v_3,v_4\}$
where $B_1$ is its parent block and $v_2$ is its control node. We have $\hat{B}_2 = (B_2\setminus B_1) = \{v_3,v_4\}$. The block structure of $\bn$ is shown in Figure~\ref{fig:blocks}.

\begin{figure}[!h]
  \begin{minipage}{0.4\textwidth}
    \centering
{\begin{tikzpicture}
[attr/.style={draw=none,fill=black!30,rounded corners}]
\draw[draw=none,fill=black!30,rounded corners,opacity=0.5] (0,-0.4) -- (-0.5,-0.4) -- (-0.5,0.4) -- (0.5,0.4) -- (0.5,-0.4) -- (0,-0.4);
\draw[draw=none,fill=black!30,rounded corners,opacity=0.5] (0,0.6) -- (-0.5,0.6) -- (-0.5,3.4) -- (0.5,3.4) -- (0.5,0.6) -- (0,0.6);
\node (t1)[attr] at (0,0) {11};
\node (t2)[attr] at (0,1) {10};
\node (t3) at (0,2) {00};
\node (t4) at (0,3) {01};
\node at (0,-0.8) {(a)};
\path
(t4) edge [->] (t3)
(t3) edge [->] (t2)
;
\end{tikzpicture}}
    \end{minipage}
  \begin{minipage}{0.5\textwidth}
    \centering
{\begin{tikzpicture}
[attr/.style={draw=none,fill=black!30,rounded corners}]
\draw[draw=none,fill=black!30,rounded corners,opacity=0.5] (1.5,0.4) -- (5.1,0.4) -- (5.1,-2.4) -- (0.9,-2.4) -- (0.9,0.4) -- (1.5,0.4);
\draw[draw=none,fill=black!30,rounded corners,opacity=0.5] (6,1.4) -- (6.6,1.4) -- (6.6,-2.4) -- (5.4,-2.4) -- (5.4,1.4) -- (6,1.4);
\node (t2)[attr] at (1.5,0) {1010};
\node (t3) at (3,0) {1011};
\node (t4) at (4.5,0) {1001};
\node (t6) at (1.5,-1) {0010};
\node (t7) at (3,-1) {0011};
\node (t8) at (4.5,-1) {0001};
\node (t10) at (1.5,-2) {0110};
\node (t11) at (3,-2) {0111};
\node (t12) at (4.5,-2) {0101};
\node (t14) at (6,1) {1101};
\node (t15) at (6,0) {1111};
\node (t16) at (6,-1) {1110};
\node (t17)[attr] at (6,-2) {1100};
\node at (3.8,-2.8) {(b)};
\path
(t3) edge [->] (t2)
(t4) edge [->] (t3)
(t6) edge [->] (t2)
(t7) edge [->] (t3)
(t7) edge [->] (t6)
(t8) edge [->] (t4)
(t8) edge [->] (t7)
(t10) edge [->] (t6)
(t11) edge [->] (t7)
(t11) edge [->] (t10)
(t12) edge [->] (t8)
(t12) edge [->] (t11)
(t15) edge [->] (t14)
(t16) edge [->] (t15)
;
\end{tikzpicture}}
    \end{minipage}
\caption{The TS of the blocks of $\bn$}
\label{fig:tsb}%
\end{figure}
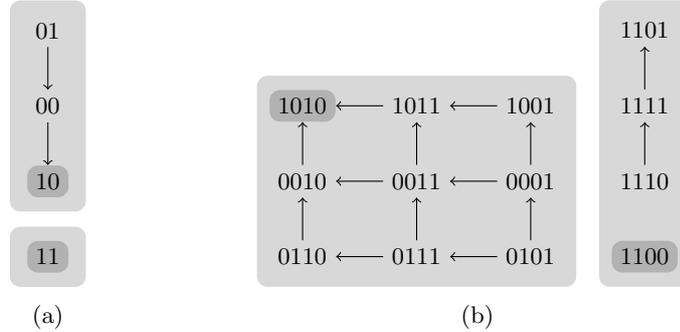

Now $A_2|_{B_1}=A_2^1=\{(11)\}$ and $A_3|_{B_1}=A_3^1=\{(10)\}$ and by Theorem~\ref{thm:pres} we know that both $A_2^1$
and $A_3^1$ are attractors of $B_1$. $\bas(A_2^1)$ and $\bas(A_3^1)$
are shown in Figure~\ref{fig:tsb}(a) where we again drop the self-loops (transitions to itself) present in all the states. The attractors $A_2^1$ and $A_3^1$ are shown in dark grey
rectangles and their their corresponding basins of attractions, $\bas(A_2^1)$
and $\bas(A_3^1)$ are shown in enclosing lighter grey regions.

Now, $\ac(B_2) = B_1\cup B_2 = V$ and $\hat{B_2} = B_2\setminus B_1=\{v_2,v_3\}$. Hence $A_2|_{\ac(B_2)}=A^2_2=A_2$ and $A_3|_{\ac(B_2)}=A^2_3=A_3$ are attractors of $\ac(B_2)$ by Theorem~\ref{thm:pres}.
We then compute $\bas(A_2^2)$ and $\bas(A_3^2)$. The attractors $A^2_2$ and $A^2_3$ are shown in Figure~\ref{fig:tsb}(b) as dark grey rectangles and their corresponding basin of attractions are
shown by enclosing areas of a lighter shade. 

We next look at the basins of the attractors $A_2^1$ and $A_3^1$ of block $B_1$ to compute the $2\times 2$ matrix $\mat^1$, shown in Table~\ref{tab:mat1a}, recording the
control needed to move to the basin of attractor $A_3^1$ from attractor $A_2^1$ and vice-versa. The elements of $\mat^1$ are subsets of $\ind(B_1)=\{1,2\}$, the indices of the vertices in
$B_1$. We similarly
look at the basins of attractions of $A_2^2$ and $A_3^2$ to compute the $2\times 2$ matrix $\mat^2$ shown in Table
\ref{tab:mat1b}. In $\mat^2$ we record the control needed to move  $A_2^2$ to the basin of $A_3^2$ and vice-versa. The elements
of $\mat^2$ are subsets of $\ind(\hat{B}_2)=\{3,4\}$, the indices of the vertices in $(B_2\setminus B_1)$.

  \begin{table}
    \centering
    \begin{minipage}{0.35\textwidth}
      \centering
    \begin{tabular}{|c||c|c|}
      \hline
           & $A^1_2$ & $A^1_3$ \\
      \hline
      \hline
      $A_2^1$ & & \{2\}\\
      \hline
      $A_3^1$ & \{1\}, \{2\}, \{1,2\} & \\
      \hline
    \end{tabular}
    \subcaption{\text{The $2\times 2$ matrix $\mat^1$}\label{tab:mat1a}}
    \end{minipage}
    \begin{minipage}{0.55\textwidth}
      \centering
      \begin{tabular}{|c||c|c|}
      \hline
           & $A^2_2$ & $A_3^2$\\
      \hline
      \hline
      $A_2^2$ & & $\emptyset, \{3\}, \{4\}, \{3,4\}$\\
      \hline
      $A_3^2$ & \{3\}, \{4\}, \{3,4\} & \\
      \hline
      \end{tabular}
      \subcaption{\text{The $2\times 2$ matrix $\mat^2$}\label{tab:mat1b}}
    \end{minipage}
    \caption{The control matrices for the blocks $B_1$ and $B_2$.}
    \label{tab:mat1}
        \vspace{-5mm}
  \end{table}

We next construct the subset lattices $\lat_1$ and $\lat_2$ corresponding to $\mat^1$ and $\mat^2$, respectively.
The elements of $\lat_1$ and $\lat_2$ are subsets of $\ind(B_1)=\{1,2\}$ and those of
$\lat_2$ are subsets of $\ind(\hat{B}_2) = \{3,4\}$. Each element $L$ of $\lat_1$ is labelled  with tuples in $(\{2,3\}\times\{2,3\})$ [corresponding to pairs of attractors in $\{A_2^1, A_3^1\}$] as:
$L$ is labelled with $(i,j)$ if and only if $L$ is an element of $\mat^1_{A_iA_j}$. Similarly, each element $L$ of $\lat_2$ is labelled with
tuples in $(\{2,3\}\times\{2,3\})$ [corresponding to pairs of attractors in $\{A_2^2, A_3^2\}$] as:
$L$ is labelled with $(i,j)$ if and only if $L$ is an element of $\mat^2_{A_iA_j}$. The lattices $\lat_1$ and $\lat_2$ with their labels are shown
in Figure~\ref{fig:lat12}.

\begin{figure}[!t]
    \centering
    \begin{minipage}{0.45\textwidth}
      \centering
      \begin{tikzpicture}
        \centering
        \node (min) at (0,0) {$\emptyset$};
        \node (a) at (-2,2) {\{1\}};
        \node [red,right] at (a.east) {\scriptsize{$(3,2)$}};
        \node [draw=none,fill=black!25,rounded corners](b) at (2,2) {\{2\}};
        \node [red,left] at (b.west) {\scriptsize{$(2,3),(3,2)$}};
        \node (max) at (0,4) {\{1,2\}};
        \node [red,right] at (max.east) {\scriptsize{$(2,3),(3,2)$}};
         \draw (min) -- (a) -- (max) -- (b) -- (min); 
      \end{tikzpicture}
      \subcaption{\text{The lattice $\lat_1$.}\label{tab:l1}}
    \end{minipage}
      \begin{minipage}{0.45\textwidth}
        \centering
        \begin{tikzpicture}
        \centering
        \node (min) at (0,0) {$\emptyset$};
        \node [red,right] at (min.east) {\scriptsize{$(2,3)$}};
        \node [draw=none,fill=black!25,rounded corners](a) at (-2,2) {\{3\}};
        \node [red,right] at (a.east) {\scriptsize{$(2,3),(3,2)$}};
        \node [draw=none,fill=black!25,rounded corners](b) at (2,2) {\{4\}};
        \node [red,right] at (b.east) {\scriptsize{$(2,3),(3,2)$}};
        \node (max) at (0,4) {\{3,4\}};
        \node [red,right] at (max.east) {\scriptsize{$(2,3),(3,2)$}};
        \draw (min) -- (a) -- (max) -- (b) -- (min); 
        \end{tikzpicture}
        \subcaption{\text{The lattice $\lat_2$.}\label{tab:l2}}
        \end{minipage}
        \caption{The subset lattices for the blocks with labels closed under subsets.}
        \label{fig:lat12subset}
  \end{figure}
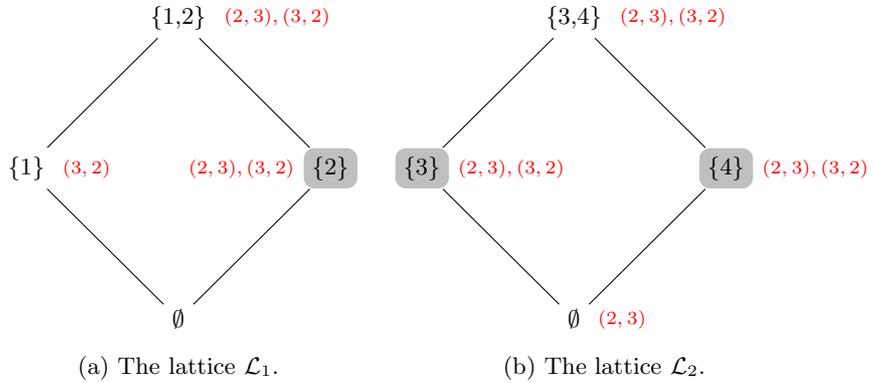

We then take the closure of the labels of the elements of $\lat_1$ and $\lat_2$ under subsets. That is, we label every element with the label of itself
and that of all its subsets. The resulting labelling of $\lat_1$ and $\lat_2$ is shown in Figure~\ref{fig:lat12subset}. We find that in $\lat_1$, $\control^1=\{2\}$ is the minimal element of $\lat_1$
whose label contains both $(2,3)$ and $(3,2)$, i.e., all the pairs of the indices of the attractors $\{A_2^1, A_3^1\}$. Similarly, in $\lat_2$, $\control^2$ is either $\{3\}$ or $\{4\}$. Each is the
minimal element of $\lat_2$ whose label contains both the pairs $(2,3)$ and $(3,2)$, i.e., all the pairs of the indices of the attractors $\{A_2^2, A_3^2\}$. Combining, we have
that $\control_\attr=\control_\bn = \control^1\cup \control^2$ is either $\{2,3\}$ or $\{2,4\}$.
  
Note that the lattice for the full TS of $\bn$, $\lat$ had $2^4=16$ elements. On the other hand, for the decomposition-based approach we computed two smaller lattices $\lat_1$ and $\lat_2$ each
of which has $2^2=4$ elements and hence the total size of the lattices is $4+4=8$. Thus, we believe that for many well-structured real-life BNs the decomposition-based approach for computing
the minimal (all-pairs and full) control might be more efficient than a global approach. This was already shown by us in~\cite{PSPM18} for the case of target control and we are currently extending
our implementation to (all-pairs and full) control to test in efficiency on various real-life BNs for biological systems.

  \begin{figure}[!t]
    \centering
    \begin{minipage}{0.45\textwidth}
      \centering
      \begin{tikzpicture}
        \centering
        \node (min) at (0,0) {$\emptyset$};
        \node (a) at (-2,2) {\{1\}};
        \node [red,right] at (a.east) {\scriptsize{$(3,2)$}};
        \node (b) at (2,2) {\{2\}};
        \node [red,left] at (b.west) {\scriptsize{$(2,3),(3,2)$}};
        \node (max) at (0,4) {\{1,2\}};
        \node [red,right] at (max.east) {\scriptsize{$(3,2)$}};
         \draw (min) -- (a) -- (max) -- (b) -- (min); 
      \end{tikzpicture}
      \subcaption{\text{The lattice $\lat_1$}\label{tab:l1}}
    \end{minipage}
      \begin{minipage}{0.45\textwidth}
        \centering
        \begin{tikzpicture}
        \centering
        \node (min) at (0,0) {$\emptyset$};
        \node [red,right] at (min.east) {\scriptsize{$(2,3)$}};
        \node (a) at (-2,2) {\{3\}};
        \node [red,right] at (a.east) {\scriptsize{$(2,3),(3,2)$}};
        \node (b) at (2,2) {\{4\}};
        \node [red,right] at (b.east) {\scriptsize{$(2,3),(3,2)$}};
        \node (max) at (0,4) {\{3,4\}};
        \node [red,right] at (max.east) {\scriptsize{$(2,3),(3,2)$}};
        \draw (min) -- (a) -- (max) -- (b) -- (min); 
        \end{tikzpicture}
        \subcaption{\text{The lattice $\lat_2$}\label{tab:l2}}
        \end{minipage}
        \caption{The subset lattices for the blocks.}
        \label{fig:lat12}
  \end{figure}
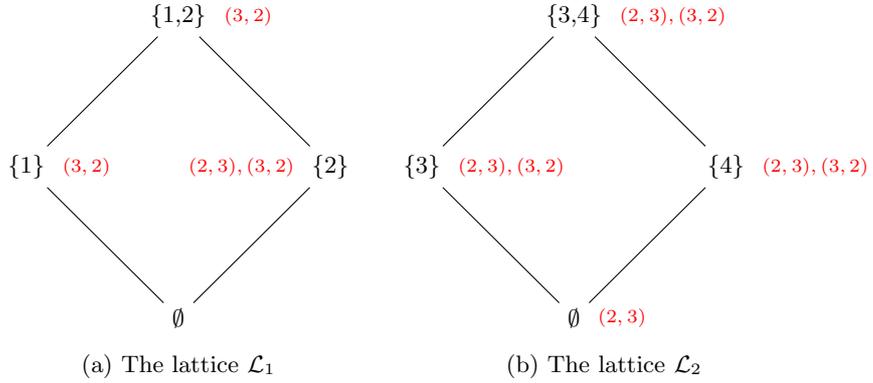

\section{Conclusion}\label{sec:conclusion}
%
In this report, we describe work-in-progress on the development of a procedure for the computation of a minimal subset of nodes
required for the existential control of a given BN. Our procedure can be applied on the entire BN in one-go or on the `blocks' of the BN locally and then
later combined to derive the global control, whereby taking advantage of the decomposition-based approach towards the problem of target control
of BNs that we have developed in~\cite{PSPM18}. We are currently implementing our procedure in software to test its efficacy and efficiency on various
real-life and random BNs. We believe that our decomposition-based approach has great potential to efficiently solve the control problem for large
real-life biological networks modelled as BNs that are modular and well-structured.

\bibliographystyle{splncs04}
\bibliography{control}
\end{document}